\documentstyle[prd,aps,preprint,epsfig,floats]{revtex}
\tightenlines

\input epsf.tex
\def\DESepsf(#1 width #2){\epsfxsize=#2 \epsfbox{#1}}
\begin{document}

\draft

%@twocolumnfalse\endcsname
\preprint{\vbox{
\hbox{UMD-PP-03-039}
}}

\title{{\Large\bf Minimal SUSY SO(10), b-$\tau$ Unification and Large
Neutrino Mixings}}
\author{\bf H. S. Goh, R.N. Mohapatra and Siew-Phang Ng }

\address{ Department of Physics, University of Maryland, College Park, MD
20742, USA}
\date{March, 2003}
\maketitle

\begin{abstract}
We show that the assumption of type II seesaw mechanism for small neutrino
masses coupled with $b-\tau$ mass unification in a minimal SUSY 
SO(10) model
leads not only to a natural understanding of large atmospheric mixing
angle ($\theta_{23}$) among neutrinos, as recently
pointed out, but also to  large solar angle ($\theta_{12}$) and a small
$\theta_{13}\equiv U_{e3}$ as required to fit observations. This is
therefore a
minimal, completely realistic grand unified model for all low energy
observations that naturally explains the diverse mixing patterns between
the quark and leptons without any additional inputs such as extra global 
symmetries. The proposed long baseline neutrino experiments
will provide a crucial test of this model since it predicts
$U_{e3}\simeq 0.16$ for the allowed range of parameters.
\end{abstract}

\vskip1.0in
\newpage

\section{Introduction}
The various neutrino oscillation experiments such as those involving solar
and atmospheric neutrinos as well as the KEK and KAMLAND experiments
that involve laboratory produced neutrinos have now produced quite
convincing evidence that neutrinos have mass and they mix among
themselves. Although the neutrinos have a great deal of similarity with
quarks as far as the weak interactions go, the oscillation results have
revealed a profound difference i.e. two of the three neutrino mixings are
very large whereas all quark mixing angles are very
small. Understanding this difference is a major challenge for
theoretical particle physics today. The problem becomes particularly
acute in models that unify quarks and leptons such as the SO(10) grand
unified models\cite{so10}, which are considered as prime candidates for
describing neutrino masses.

Some of the reasons that make  SO(10) models so attractive as grand
unification theories of nature are the following: (i) in SO(10) model,
 all fermions can be part of a single spinor representation; (ii) it
contains
the left-right symmetric unification group $SU(2)_L\times SU(2)_R\times
U(1)_{B-L}\times SU(3)_c$\cite{lr}
which provides a more satisfactory way to understand the origin of parity
violation in Nature and (iii) finally, perhaps the single most important
reason is the
natural understanding of small neutrino masses via the seesaw
mechanism\cite{seesaw} since the single spinor representation discussed
above that contains all the standard model fermions also contains the
right handed neutrino, needed in implementing the seesaw mechanism.

A closer look at the details of neutrino oscillation data in fact provides
one more compelling reason for the SO(10) model: in order
to understand the atmospheric neutrino data, we need
 the heaviest neutrino mass to be larger than $0.05$ eV. The
 seesaw formula i.e. $m_{\nu}\sim \frac{m^2_D}{M_R}$ (where $m_D$ is
the neutrino Dirac mass and $M_R$ is the mass of the right handed neutrino
contributing to the neutrino mass needed to understand the atmospheric
data) then tells us that there must be one right handed $M_R \leq
10^{15}$ GeV. This value is considerably smaller
 than the Planck mass and therefore one is faced with a new
hierarchy problem similar to
the corresponding problem of the standard model. However, it was pointed
out long ago\cite{marshak} that the Majorana mass of the RH neutrino owes
its origin to the breaking of local B-L symmetry which implies that
$M_R\simeq M_{B-L}$. Local B-L symmetry therefore provides a natural way
to understand
the smallness of the RH neutrino mass compared to $M_{P\ell}$. What is
very interesting is that SO(10) group also contains the local B-L as a
subgroup.

SO(10) model, despite its attractiveness for understanding the overall
scale of neutrino masses, runs into a potential trouble in providing an
understanding of the observed mixings. The problem
arises from the fact that SO(10) also contains the quark-lepton
unification $SU(4)_c$ group of Pati and Salam, which in the simplest
approximation leads to equal quark and lepton mixing angles and one needs
to make further assumptions to get a handle on the mixings\cite{many}.
 An obvious conceptual problem is that if one of these
models is ruled out by data, one would not be able to tell whether it is
the SO(10) unification which is ``at fault'' or it is one of the
assumptions used to derive neutrino mixings.

A different approach to this issue was
taken in ref.\cite{babu}. The idea was to avoid the use of any symmetries
beyond the gauge symmetry, in this case SO(10) and
use the minimal set of Higgs fields that can break the group down to the
standard model and give mass to the fermions. It turns out that if we
choose Higgs fields in {\bf 10} and ${\bf \overline{126}}$, the Yukawa
superpotential contains enough parameters to fit the observed
fermion masses and mixings of the standard model.  It was observed in
ref.\cite{babu}, that in this model the neutrino masses and mixings are
completely predicted upto an overall scale, when one uses
the seesaw mechanism which is part of the SO(10) model\footnote{This is
to be contrasted with the SU(5) case where
the minimal Higgs set needed to break the gauge symmetry i.e. {\bf 5}+{\bf
24} Higgses lead to the mass relation $m_e/m_{\mu}~=~m_d/m_s$
that is in contradiction with observations.}. To break the gauge symmetry
fully, two additional Higgs multiplets belonging to {\bf 45}+{\bf
54} are included. They do not contribute to the fermion masses, leaving
the conclusions on neutrino masses unchanged.

An additional appeal of breaking B-L symmetry of  SUSY SO(10) by an {\bf
126}, as opposed to by {\bf 16} Higgs, is
that it automatically leaves R-parity as an exact aymmetry and thereby
explains, why neutralino is a stable dark matter\cite{rabi,gora}. This is
because the
submultiplet of {\bf 126} that breaks B-L carries B-L =2. Therefore
R-parity (defined by $Rp=(-1)^{3(B-L)+2S}$) quantum number of this field
is even and therefore, its vev leaves R-parity unbroken. In contrast in
models where B-L is broken
by a {\bf 16}-plet of Higgs, the B-L symmetry is broken by one unit and 
without any additional symmetries (e.g. matter parity), neutralino is
unstable and cannot therefore serve as a dark matter. Of course, if a
fundamental theory  e.g. a superstring theory that led to an SO(10) model
with appropriate additional symmetries that guarantee the stability of
neutralino was known, then the above objection to an {\bf 16} Higgs would
not apply.

 The initial analyses of neutrino mixings within the framework of 
Ref.\cite{babu} used the simple seesaw formula (type I seesaw) and are
now in disagreement with data. In subsequent
papers\cite{lav,lee,brahma,takasugi,fukuyama}, this idea has been analysed
( in some cases by including more than one {\bf 10} Higgses) to see how
close one can come close to the observed neutrino parameters. The
conclusion now appears to be that one needs CP violating phases to achieve
this goal, as noted in \cite{fukuyama}.

A way out of this problem is to use the type II seesaw
mechanism, as was initially done in \cite{brahma}, where an induced
triplet vev is added to the usual type I seesaw formula arising from the
RH neutrino intermediate state. In models which have asymptotic parity
symmetry such as left-right or SO(10) models, type II seesaw arises if
both parity and B-L symmetry are broken at the same scale.  

A very interesting point about this approach has been noted in a recent
paper\cite{bajc}, where it has been shown that if we restrict ourselves
to the 2-3 sector of the model and use the type II seesaw mechanism, then
the $b-\tau$ unification of supersymmetric grand unified theories leads to
a neutrino Majorana mass matrix which explains the large
$\nu_\mu-\nu_\tau$
mixing angle needed to understand atmospheric neutrino data. The important
point is that no symmetries are needed to get this result. 

To see how this result is very generic to SO(10) models with {\bf
126} vev, note that: (i) in SO(10) model, the neutrino mass
matrix is given by the type II seesaw formula, with two
contributions: one coming from the righthanded neutrino intermediate
state and another coming from an induced triplet vev generic to these 
models, as already mentioned; and (ii) that for certain range of
parameters, the induced  
triplet vev term can dominate the neutrino mass matrix. As was shown in
\cite{brahma}, under these circumstances, one gets a sumrule
\begin{eqnarray}
{\cal M}_{\nu}~=~ a ({\cal M}_\ell-{\cal M}_d)
\label{key}
\end{eqnarray}
In \cite{bajc}, it was observed that since this relation is valid at the
seesaw scale, one must use the extrapolated quark and lepton masses in the
formula. The fact that at or near the GUT scale $m_b/m_\tau~\simeq 1-1.2$
depending on the value of $tan\beta$, implies that the 3-3 element of the
${\cal M}_\nu$ which is proportional to $m_b-m_\tau$ is of the same
order as the off diagonal elements of the ${\cal M}_{nu}$
 in the 2-3 subsector leading top the largeness of the atmospheric
mixing angle without any further assumptions. 

It is however essential to do a complete three generation
analysis of this model if this important observation is to lead to a
realistic SO(10) model for understanding all neutrino mixings. In fact,
since the model has no free parameters, it is a priori not obvious that
within
this framework one would simultaneously get a large solar mixing angle and
a small $\theta_{13}\equiv U_{e3}$ as well as the correct value for the
ratio $\Delta
m^2_{\odot}/\Delta m^2_A$. It is the goal of this paper to analyze this
question.

We find that for a
 narrow range of the quark masses, the model does indeed lead to the
correct mass
difference-squares as well as large mixing angles $\theta_{12}$ and
$\theta_{23}$ needed to understand the solar and atmospheric neutrino
oscillations while at the same time keeping $U_{e3}\leq 0.16$ as
required by the reactor data. We find that for this entire range, the
reactor angle $U_{e3}\simeq 0.16$ providing a clear way to test the
model in the proposed long base line neutrino experiments.

This paper is organized as follows: in sec. II, we review the basic
equations of the model for the supersymmetric case; in sec. III, we
explain the conditions under which the induced triplet vev dominates
neutrino masses; in sec. IV, we
describe our method for solving the equations to predict the neutrino
parameters. In sec. V, we give our results for the neutrino masses and
the mixing angles. In sec. VI, we give our conclusions. In this paper,
we ignore the CP violating phases.

\section{The mass sumrules for minimal SO(10)}
We consider supersymmetric SO(10) group with the Higgs fields belonging to
the representations {\bf 45}+{\bf 54} for breaking SO(10) group down
to the left-right symmetric group  $SU(2)_L\times SU(2)_R\times
U(1)_{B-L}\times SU(3)_c$ and the minimal Higgs set {\bf 10}+{\bf
126}+${\bf \overline{126}}$ that couple to matter and also break the
$SU(2)_L\times SU(2)_R\times U(1)_{B-L}\times SU(3)_c$ group down to
$SU(3)_c\times U(1)_{em}$. It
is the latter set i.e. ${\bf 10}\oplus{\overline{\bf 126}}$ which is
crucial to
our discussion of fermion masses. The first stage
of the symmetry breaking therefore could have been accomplished by
alternative Higgs multiplets e.g. by {\bf 210} of Higgs without
effecting our results. As has been noted
earlier\cite{babu,lee}, the set {\bf 10}+${\bf \overline{126}}$ which
couple
to matter contain two pairs of MSSM Higgs doublets belonging to
(2,2,1) and (2,2,15) submultiplets (under $SU(2)_L\times SU(2)_R\times
SU(4)_c$ subgroup of SO(10)). We denote the two pairs by $\phi_{u,d}$
and $\Delta_{u,d}$. At the GUT scale, by some doublet-triplet
splitting mechanism these two pairs reduce to the MSSM Higgs pair
$(H_u,H_d)$, which can be expressed in terms of the $\phi$ and $\Delta$ as
follows:
\begin{eqnarray}
H_u~=~cos\alpha_u \phi_u + sin\alpha_u \Delta_u \\ \nonumber
H_d~=~cos\alpha_d \phi_d + sin\alpha_d \Delta_d
\end{eqnarray}
The details of the doublet-triplet splitting mechanism that leads to the
above equation are not relevant for what follows and we do not discuss it
here. As in the case of MSSM, we will assume that the Higgs doublets
$H_{u,d}$ have the vevs $<H^0_u>=v sin\beta$ and $<H^0_d>=v cos\beta$.

In orders to discuss fermion masses in this model, we start with the
SO(10) invariant superpotential giving the Yukawa couplings of the {\bf
16} dimensional matter spinor $\psi_i$ (where $i,j$ denote generations) 
with the Higgs fields $H_{10}\equiv
{\bf 10}$ and $\Delta\equiv {\bf \overline{126}}$.
\begin{eqnarray}
{W}_Y~=~ h_{ij}\psi_i\psi_j H_{10} + f_{ij} \psi_i\psi_j\Delta
\end{eqnarray}
SO(10) invariance implies that $h$ and $f$ are symmetric matrices.
We ignore the small effects coming from the higher dimensional operators.
Below the B-L breaking (seesaw) scale, we can write the superpotential
terms for the charged fermion Yukawa couplings as:
\begin{eqnarray}
W_0~=~h_u QH_uu^c + h_d QH_d d^c + h_eLH_d e^c + \mu H_uH_d
\end{eqnarray}
where
\begin{eqnarray}
h_u~=~hcos\alpha_u + fsin\alpha_u\\ \nonumber
h_d~=~hcos\alpha_d + fsin\alpha_d\\ \nonumber
h_e~=~hcos\alpha_d -3 fsin\alpha_d
\end{eqnarray}
In general $\alpha_u\neq \alpha_d$ and this difference is responsible for 
nonzero CKM mixing
angles. In terms of the GUT scale Yukawa couplings, one can write the
fermion mass matrices at the seesaw scale as:
\begin{eqnarray}
M_u~=~(\bar{h} + \bar{f} )\\ \nonumber
M_d~=~(\bar{h}r_1 + \bar{f}r_2 )\\ \nonumber
M_e~=~(\bar{h}r_1 -3r_2 \bar{f} )\\ \nonumber
M_{\nu^D}~=~(\bar{h} -3 \bar{f} )
\end{eqnarray}
where
\begin{eqnarray}
\bar{h}=h cos\alpha_u sin\beta\\ \nonumber
\bar{f}= f sin\alpha_u sin\beta\\ \nonumber
r_1~=~\frac{cos\alpha_d}{cos\alpha_u}cot\beta\\ \nonumber
r_2~=~\frac{sin\alpha_d}{sin\alpha_u}cot\beta
\end{eqnarray}
To count the number of parameters describing the fermion sector, we first
 ignore CP phases and choose a basis where $\bar{h}$ is diagonal. Since
$\bar{f}$ is symmetric, we have a total of nine parameters from the
couplings and including $\alpha_{u,d}$ and $\beta$ gives us a total of
twelve parameters. All these parameters can be determined by fitting the
the six quark masses, three lepton masses and three CKM angles. This
enables a complete determination of the neutrino masses upto an overall
scale related to the B-L symmetry breaking and the three
mixing angles. The model is therefore completely predictive in the
neutrino sector.

In order to determine the neutrino masses and mixings, one uses the seesaw
mechanism. As noted in the introduction, most previous works on this model
except the works in
Ref.\cite{brahma,bajc} used the type one seesaw mechanism where the
neutrino mass matrix is given by the formula:
\begin{eqnarray}
{\cal M}_\nu~=~-M_{\nu^D}M^{-1}_{N_R}M^T_{\nu^D}
\end{eqnarray}
where $M_{N_R}= f v_{B-L}$. On the other hand it is well known that in
asymptotically parity
conserving theories including SO(10), the true seesaw
formula\cite{seesaw2} (called type II
here and in literature) has a second term which arises from an
induced
$SU(2)_L$ triplet vev or from higher dimensional terms involving left
doublets:
 \begin{eqnarray}
{\cal M}_\nu~=~\bar{f}\sigma_L-M_{\nu^D}M^{-1}_{N_R}M^T_{\nu^D}
\end{eqnarray}
 where $\sigma_L~=~\lambda \frac{v^2}{v_{B-L}}$, where $v$ is the
$SU(2)_L$ breaking scale and $\lambda$ is a combination of parameters 
in the Higgs potential. The type II seesaw
formula has two new parameters $\sigma_L$ and $v_{B-L}$ instead of one
in the type I case; but for different ranges of initial
parameters in Higgs potential, either the
first or the second term can be made to dominate\cite{chang}. We will work
within the assumption that it is the first term that dominates the seesaw
formula\cite{bajc}. This situation can arise when $\sigma_L\gg
v^2/v_{B-L}$ (or $\lambda$ which is a ratio of scalar coupling parameters
is much larger than one). We elaborate the circumstances when this happens
in sec. III.

It was noted in several papers\cite{babu,lee,brahma,takasugi} that if one
uses type I seesaw, this model cannot produce two large neutrino mixing
angles if CP phases are ignored. It
has been shown recently\cite{fukuyama} that once the CP phases are
included, one can get bi-large neutrino mixing pattern. But here we search
for two large mixing angle solutions without invoking the CP phase. As
noted in
\cite{bajc}, if one uses type II seesaw and assumes further that the first
term dominates, large atmospheric mixing angle follows naturally as a
consequence of b-$\tau$ unification. Whether this also simultaneously
yields a large solar mixing angle and a small $U_{e3}$ remained an open
question. In this present paper we present
a detailed analysis of this idea in a full three generation model and show
that for a very narrow range of quark masses, this model also predicts
large solar mixing angle
as well as the correct solar mass splitting together with a small
$U_{e3}$.
Since the model has hierarchical mass pattern, the
atmospheric mass difference-squared is related to the overall scale of the
triplet vev in the type II seesaw formula and cannot be predicted.

\section{Dominance of induced triplet vev}
In this section, we would like to discuss the parameter range where the
induced triplet vev term dominates the neutrino mass matrix. To elucidate
this let us discuss the origin of the triplet vev in our minimal
SO(10) model. First we note the decomposition of the ${\bf \bar{126}}$
under the group $SU(2)_L\times SU(2)_R\times SU(4)_c$:
\begin{eqnarray}
{\bf \overline{126}}~=~(1,1,6)\oplus (2,2,15)\oplus (3,1,\bar{10}) \oplus
(1,3,10)
\end{eqnarray}
The $SU(2)_L$ triplet that contributes to the type II seesaw formula is
contained in the multiplet $\Delta_L\equiv (3,1,\overline{10})$ and it
couples
to the left
handed multiplet $\psi \equiv (2,1,4)$ of the {\bf 16} dimensional
SO(10) spinor that contains the matter fermions 
i.e. $\psi_L\psi_L\Delta_L$. On the other hand the mass of the RH
neutrinos comes from the coupling of $\Delta_R\equiv (1,3, 10)$
submultiplet of ${\bf \overline{126}}$ to the right handed fermion
multiplet
$\psi^c_L\equiv (1,2,\bar{4})$ i.e. $\psi^c_L\psi^c_L\Delta_R$.

The vev of the neutral member of $\Delta_R$ breaks the B-L symmetry and
gives mass to the RH neutrinos and generates the second term in the type
II seesaw formula. To see how the $\Delta^0_L$ vev arises, note that the
general superpotential of the model contains terms of type $\lambda_1{\bf
126}^2\cdot {\bf 54}$ and $\lambda_2{\bf 10}\cdot {\bf 10}\cdot {\bf
54}$. In the Higgs potential, this
generates a term (from $|F_{\bf 54}|^2$) of the form ${\bf 10\cdot 10\cdot
\overline{126}\cdot \overline{126}}$. In this term, there is a term of the
form 
$\phi(2,2,1)^2\Delta_L(3,1,10)\Delta_R(1,3,\bar{10})$ with a coefficient
$\lambda_1\lambda_2$. Furthermore, in the Higgs potential, there is a mass
term for $\Delta_L(3,1,10)$ of the form $\mu^2_{\Delta}+\lambda_3 v^2_U$,
where $v_U$ is the GUT scale. On minimizing the potential, these two terms
lead to a vev for the $SU(2)_L$ triplet $\sigma_L\equiv <\Delta^0_L>\simeq
\frac{\lambda_1\lambda_2 v^2_{wk} v_{B-L}}{\mu^2_{\Delta}+\lambda_3
v^2_U}$. It is now clear that if we choose $\lambda_3$ such that 
$\mu^2_{\Delta}+\lambda_3 v^2_U \ll v^2_{B-L}$, then the entries
in the second matrix in the type II seesaw formula can much smaller than
$\sigma_L$  and Eq.\ref{key} holds. We will work in the domain of the
parameter space where this happens.

If the triplet vev contribution to the neutrino mass matrix dominates in
the type II seesaw formula, then Equation (6) can be used to derive the
sumrule
\begin{eqnarray}
{\cal M}_{\nu}~=~a(M_{\ell}-M_d)
\end{eqnarray}
Using this equation in second and third generation sector, one can
understand the results of \cite{bajc} in a heuristic manner as follows.
The known hierarchical structure of quark and lepton masses as well as
the known small mixings for quarks suggest that $M_{\ell,d}$ have the
following pattern
\begin{eqnarray}
M_{\ell}~\simeq~m_\tau\pmatrix{\epsilon_{\ell,1}&\epsilon_{\ell,2}\cr
\epsilon_{\ell,2} & 1} 
\end{eqnarray}
and\begin{eqnarray}
M_{d}~\simeq~m_b\pmatrix{\epsilon_{d,1}&\epsilon_{d,2}\cr
\epsilon_{d,2} & 1}
\end{eqnarray}
where $\epsilon_{\ell,d;i}\ll 1$. It is then clear that if there is
approximate $b-\tau$ unification as it appears to be the case if the
theory below the
B-L breaking scale is MSSM, then in $M_\ell-M_d$ matrix, there is a high
degree of cancellation in the $33$ entry making this entry comparable to
all the other entries in this matrix. The atmospheric mixing angle which
is given by $tan 2\theta_A\simeq
(m_\tau\epsilon_{\ell,2}-m_b\epsilon_{d,2})/(m_b-m_\tau+m_b\epsilon_{d,1}-
m_l\epsilon_{\ell,1})$ becomes very large at the B-L breaking
scale. Since
renormalization played an important role in obtaining this result, one
must ask what happens to the neutrino mixings once they are extrapolated
to the weak scale\cite{RGE}. It is well known\cite{RGE} that for the case
of normal
hierarchy for neutrino masses as is the case here, the MSSM RGE's do not
change the mixing angles very much and the seesaw scale result persists at
the weak scale with only minor changes.

 \section{Details of Calculation}
In this section, we outline our method for determining the neutrino mixing
parameters. For this purpose, we first note that the matrices $\bar{h}$
and $\bar{f}$ in Eq. (6) can be eliminated in terms the mass matrices
$M_{u,d}$ so that we have a sumrule involving the three mass matrices
$M_{u,d,\ell}$. Before giving the sum rule, we note that we will work in a
basis where $M_d$ is
diagonal and is given by $M_u=V^T\cdot M_u^D \cdot V$ (where $M_u^D$
is the diagonal mass matrix of up type quark and $V$ is the CKM mixing
matrix). This can be done without any loss of generality. 
We also introduce a new set of matrices $\tilde{M}_{l,
u,d}$  where $\tilde{M}\equiv\frac{M}{m_{3}}$, $m_3$ being the third
family mass for the corresponding flavor.
The sumrule for charged lepton matrices is 
given by:
\begin{equation}\label{main}
    k \tilde{M}_l=r \tilde{M}_d+\tilde{M}_u
\end{equation}
where $k$ and $r$ are functions of $r_{1,2}$ (given in sec. II) and
fermion masses as
follows:
\begin{eqnarray}
    k=\frac{r_2-r_1}{4r_1r_2}\frac{m_\tau}{m_t} \\
    r=-\frac{r_2+3r_1}{4r_1r_2}\frac{m_b}{m_t}
\end{eqnarray}
\begin{equation}\label{mnu}
    {\cal M}_\nu=~a(\frac{m_b}{m_\tau}\tilde{M}_d-\tilde{M}_l)
\end{equation}
 These relations are valid at the
B-L breaking scale $v_{B-L}$. The advantage of working with $\tilde{M}$
rather tha $M$ is that the $33$ elements of all $\tilde{M}_{l, u,d}$
matrices are either one or of order one; so we expect solutions for $k$
and $r$ also of order
one. Furthermore since the formula for ${\cal M}_{\nu}$ involves only
$M_{\ell}$ and $M_d$, $b-\tau$ unification helps to see the cancellation
in  the $33$ element of ${\cal M}_\nu$ somewhat more easily. At the same
time the $23$ element
of ${\cal M}_\nu$ receives only one contribution from $M_{\ell}$ since in
our basis $M_d$ is diagonal. These two results lead
to atmospheric mixing angle being large\cite{bajc}. 

To carry out the calculations, we have to solve for the two unknowns $k$
and $r$ using the low energy
inputs from the quark and charged lepton sectors. To obtain a perturbative
estimate of these parameters, we
decompose  $r \tilde{M}_d+\tilde{M}_u$ as:
\begin{equation}\label{decom}
\left(%
\begin{array}{ccc}
  x & 0 & 0 \\
      0 & y & \epsilon_2 \\
      0 & \epsilon_2 & z \\
\end{array}%
\right)
   +\left(%
\begin{array}{ccc}
  0 & \epsilon_1 & a \\
      \epsilon_1 & 0 & 0 \\
      a & 0 & 0 \\
\end{array}%
\right)\equiv r\left(%
\begin{array}{ccc}
  d & 0 & 0 \\
      0 & s & 0\\
      0 & 0 & 1 \\
\end{array}%
\right)+\left(%
\begin{array}{ccc}
  u & \epsilon_1 & a \\
      \epsilon_1 & c & \epsilon_2 \\
      a & \epsilon_2 & 1 \\
\end{array}%
\right)
\end{equation}
where $\epsilon_i, a \ll 1$ as are $x$ and $y$.
In this analytical approach, our procedure will be to
 find the eigenvalues of (\ref{decom}) by perturbation
method and match them to the known leptonic masses at the B-L scale. The
advantage of this decomposition is that it allows a nice perturbative
determination of the eigenvalues analytically without having to resort
to immediate numerical analysis. We will compare our results with the
numerical evaluation using Mathematica.

The $i^{th}$
eigenvalue $\lambda_i=\lambda_i^{(0)}+\lambda_i^{(2)}$ is found to be
\begin{eqnarray}
    \lambda_1^{(0)}=x \\ \nonumber
    \lambda_2^{(0)}=\frac{y+z-\sqrt{(z-y)^2+4\epsilon_2^2}}{2} \\ \nonumber
     \lambda_3^{(0)}=\frac{y+z+\sqrt{(z-y)^2+4\epsilon_2^2}}{2}\sim
z+\frac{\epsilon_2^2}{z}+z O(10^{-2})
    \\\nonumber
     \lambda_2^{(2)}\simeq
\frac{(z\epsilon_1-a\epsilon_2)^2}{\lambda_2^{(0)} z^2} \simeq
    O(10^{-2})\lambda_2^{(0)}\\ \nonumber
    \lambda_3^{(2)}\simeq \frac{a^2}{\lambda_3^{(0)}} \simeq
    O(10^{-2})\lambda_3^{(0)}\\\nonumber
    \lambda_1^{(2)}=-(\lambda_1^{(2)}+\lambda_3^{(2)})
\end{eqnarray}
We consider only cases where $y\simeq 10^{-2}$ and $z>0.1$. Within
this regime, the unperturbed 2nd and 3rd lepton masses are accurate up
to a few \%. However, the higher order electron mass correction is
big and so the perturbation formula breaks down for this case. We
therefore use the
perturbation technique for the second and third generation masses but
use the determinant to
find that for the first generation.
As mentioned, we will check
the validity of perturbation result using numerical methods.

Taking determinant of the above equation \ref{decom}, we find that the 
three charged lepton
masses are related as follows:
\begin{eqnarray}\label{pert1}
     k^3 \tilde{m}_e \tilde{m}_\mu
=xyz-x\epsilon_2^2-ya^2-z\epsilon_1^2+2a\epsilon_1\epsilon_2\\
    k \tilde{m}_\mu=\lambda_2 \simeq \lambda_2^{(0)} \\
    k =\lambda_3 \simeq \lambda_3^{(0)} \simeq z+ \frac{\epsilon_2^2}{z}
\end{eqnarray}
We now solve the above equation by substituting
$x$,$y$,$z$,$a$,$\epsilon_1$,$\epsilon_2$ with the corresponding
elements in the matrix $r M_d + M_u$. From eq.(\ref{pert1}), and
find 
\begin{eqnarray}
    k (1+\tilde{m}_\mu )=y+z \\
    k =z+ \frac{\epsilon_2^2}{z}
\end{eqnarray}
Since Eq.\ref{decom} tells us that $z=1+r$ and $y=rs + c$, we can use the
above two equations to determine the parameters $k$ and $r$, which we can
then use to find neutrino masses and mixings.
We find $k$ and $r$ to be
\begin{eqnarray}\label{r}
   r=\frac{(s+c-2\tilde{m}_\mu )\pm
\sqrt{(s-c)^2-4(\tilde{m}_\mu-s)
(1+\tilde{m}_\mu)\epsilon_2^2}}{2(\tilde{m}_\mu-s)}\\\nonumber
   k=\frac{(1+s)r+1+c}{1+\tilde{m}_\mu}\\ \nonumber
\end{eqnarray}
and a consistency relation for the d-quark mass
\begin{eqnarray}
    d=\frac{k^3\tilde{m}_e\tilde{m}_\mu
 +z\epsilon_1^2+ya^2-2a\epsilon_1\epsilon_2-u(yz-\epsilon_2^2)}{r 
(yz-\epsilon_2^2)}
\end{eqnarray}
In order to get a rough feeling for the way the maximal neutrino mixings
arise, let us diagonalize the charged lepton mass matrix given in
Eq. \ref{main} and write the neutrino mass matrix in this basis:
 \begin{equation}
    {\cal M}_\nu=~a(\frac{m_b}{m_\tau}U_l^\dagger\tilde{M}_d
U_\ell-\tilde{M}_l^D)
\end{equation}
Where $\tilde{M}_l^D$ is the diagonal charged lepton mass matrix
with $\tau$ mass is $1$. $U_l$ is the rotation matrix diagonalize
charged lepton mass. $U_l$ can be written approximately as
\begin{equation}
    U_l\simeq\left(%
\begin{array}{ccc}
  1 & \delta_1 & \delta_2 \\
  \Delta_1 & \cos \phi & \sin \phi \\
  \Delta_2 & -\sin \phi & \cos \phi \\
\end{array}%
\right),
\end{equation}
where
\begin{equation}
    \tan \phi = \frac{\epsilon_2}{z-\lambda_2^{(0)}}.
\end{equation}
The parameters $\delta_i$ and $\Delta_i$ are given to lowest order in
perturbation theory by
\begin{eqnarray}
    \delta_1 = \frac{\epsilon_1 \cos \phi -a \sin\phi}{k\tilde{m}_\mu-x}
\\
    \nonumber
    \delta_2 = \frac{\epsilon_1 \sin \phi +a
    \cos\phi}{k-x}\\\nonumber
 \Delta_1 = -\delta_1\cos \phi-\delta_2 \sin \phi\\\nonumber
    \Delta_2 = \delta_1 \sin \phi-\delta_2\cos \phi
\end{eqnarray}
Using these parameters and neglecting small terms due to
$\delta_1$ and $\delta_2$ multiplying light quark masses, we find that
\begin{equation}\label{mnu1}
    M_\nu\simeq\left(%
\begin{array}{ccc}
  m_d-m_e+m_s\Delta_1^2 + m_b \Delta_2^2 & m_s
  \Delta_1 \cos\phi-m_b\Delta_2\sin\phi & m_s
  \Delta_1 \sin\phi+m_b\Delta_2\cos\phi \\
m_s\Delta_1 \cos\phi-m_b\Delta_2\sin\phi & m_s
-m_\mu+m_b\sin^2\phi & -m_b\sin\phi  \\
  m_s\Delta_1 \sin\phi+m_b\Delta_2\cos\phi & -m_b\sin\phi  &
   -m_b\sin^2\phi+m_b-m_\tau \\
\end{array}%
\right)
\end{equation}
We now find the following analytic expression for the
atmospheric mixing angle 
from Eq. \ref{mnu1} to leading order ignoring small terms to be:
\begin{eqnarray}
   \tan\theta_A \simeq\frac{2}{q+\sqrt{q^2+4}}\\ \nonumber
   q =\frac{2m_b\sin^2\phi+(m_\tau-m_b)+(m_s-m_\mu)}{m_b\sin\phi}
\end{eqnarray}
For $|q| \leq 1$, we get $\sin^22\theta_A \geq 0.8$. We see that $b-\tau$
unification i.e. $m_b\simeq m_{\tau}$ and $m_b \sin\phi \simeq
(m_b-m_\tau)$ are important to get a large
$\theta_A$. Also we need to have $m_s <0$ and $m_{\mu} > 0$.

\bigskip

\section{Predictions for Neutrino masses and mixings}

In order to obtain the predictions for neutrino masses and mixings in
our model, we
will need the values of quark masses and mixings at the seesaw scale.
Experiments determine these input parameters near the GeV scale and they
need to be extrapolated to the B-L scale which is
near $10^{15}$-$10^{16}$ GeV where our Equation (6) is valid.
Taking the values for the quark masses and mixings at the B-L scale 
we can determine $k$ and $r$ approximately. We will use this
determination of $k$ and $r$ to solve for neutrino masses and mixings
using the relation in Eq.\ref{mnu}. We will also compare our results with
a direct
numerical scan of the Eq. \ref{main} i.e. not using perturbation method to
obtain $k$ and $r$. Results obtained by both methods are in agreement.

 In our
model, the theory below the B-L
breaking scale is the MSSM whose effect on fermion mass extrapolation is a
well studied problem\cite{parida}. We will use the two loop analysis in
the paper by Das and Parida\cite{parida} in our analysis. Our strategy
will be to take the
values of the quark masses at the scale $v_{B-L}\simeq 10^{16}$ GeV given
in \cite{parida}.  In Table I, we give the input values of masses and
mixings for values of the MSSM parameter $tan \beta = 10$ and $55$.

\bigskip
\begin{center}
\begin{tabular}{|c||c||c|}\hline
input observable & $tan\beta=10$ & $tan\beta=55$ \\ \hline
$m_u$ (MeV) & $0.72^{+0.13}_{-0.14}$ &
$0.72^{+0.12}_{-0.14}$\\\hline
$m_c$(MeV) & $210.32^{+19.00}_{-21.22}$ &
$210.50^{+15.10}_{-21.15}$ \\ \hline
$m_t$(GeV) & $82.43^{+30.26}_{-14.76}$ &
$95.14^{+69.28}_{-20.65}$ \\\hline
$m_d$ (MeV) & $1.50^{+0.42}_{-0.23}$ &$1.49^{+0.41}_{-0.22}$
\\\hline
$m_s$ (MeV) & $29.94^{+4.30}_{-4.54}$ &$29.81^{+4.17}_{-4.49}$
\\\hline
$m_b$ (GeV) & $1.06^{+0.14}_{-0.08}$ &$1.41^{+0.48}_{-0.19}$
\\\hline
$m_e$ (MeV) & $0.3585$ & $0.3565$ \\\hline
$m_{\mu}$(MeV) & $75.6715^{+0.0578}_{-0.0501}$ &
$75.2938^{+1912}_{-0.0515}$\\\hline
$m_{\tau}$ (GeV) & $1.2922^{+0.0013}_{-0.0012}$ &
$1.6292^{+0.0443}_{-0.0294}$\\ \hline
\end{tabular}
\end{center}
\bigskip

\noindent{{\bf Table I:} The extrapolated values of quark and lepton
masses at the GUT scale from the last reference in \cite{parida}. We have 
kept the errors to only two significant figures in the quark masses.}

For the mixing angles at GUT scale, we take:
\begin{eqnarray}
V_{CKM}~=~\pmatrix{0.974836& 0.222899& -0.00319129\cr -0.222638 &
0.974217 &  0.0365224\cr 0.0112498& -0.0348928& 0.999328}
\end{eqnarray}
In the first perturbative method, we use the above input values to obtain
$k$ and $r$ using Eq. \ref{r} and search for values around them that give
a good fit to charged lepton masses and then use them in Eq.\ref{mnu} to
derive
the neutrino masses and the three mixing angles: $sin^22\theta_{\odot}$,
$sin^22\theta_A$ and $U_{e3}$. The best fit range for $k,r$ are $-.78 \leq
r\leq -.74$ and $0.23\leq k \leq .26$. We also do a direct numerical
solution. Both the results are in agreement. (We ignore CP
violation in this work.)

Note that the sign of a fermion is not physical, which
leads to several choices for the sign of fermion masses that we have put 
into our search for solutions. The solutions we present
here correspond to $m_{e,\mu,\tau,b,t} > 0$ and $m_{c,d,s}< 0$ upto
an overall sign. 

Our results are displayed in Fig. 1-3 for the case of the
supersymmetry parameter $tan\beta~=~10$. In these figures, we have
restricted 
ourselves to the range of quark masses for which the atmospheric mixing
angle $sin^22\theta_A \geq 0.8$. (For presently preferred range of values 
of $sin^22\theta_A$ from experiments, see \cite{kan}).
 We then present the predictions for
$sin^22\theta_{\odot}$, $\Delta m^2_{\odot}$ and $U_{e3}$ for the allowed
range $sin^22\theta_A$ in Fig.1, 2 and 3 respectively.
 The spread in the predictions come
from uncertainties in the $s,c$ and the $b$-quark masses. Note two
important predictions: (i) $sin^22\theta_{\odot}\geq 0.91$ and $U_{e3}\sim
\pm 0.16$. The present allowed range for the solar mixing angle is
$0.7 \leq sin^22\theta_{\odot} \leq 0.99$ at 3$\sigma$
level\cite{kan,bahcall}. 
The solutions for the neutrino mixing angles are 
sensitive to the $b$ quark mass.

It is important to note that this model predicts
the $U_{e3}$ value very close to the present experimentally allowed upper
limit and can therefore be tested in the planned long base line
experiments which are expected to probe
$U_{e3}$ down to the level of $\sim 0.05$\cite{numi,JHF}. Our model would
also
prefer a value of the $sin^22\theta_A$ below 0.9, which can also be used
to test the model. For instance, the JHF-Kamioka neutrino
experiment\cite{JHF} is
projecting a possible accuracy in the measurement of $sin^22\theta_A$ down
to the level of $0.01$ and can provide a test of this model.

\begin{figure}[h!]
\begin{center}
\epsfxsize15cm\epsffile{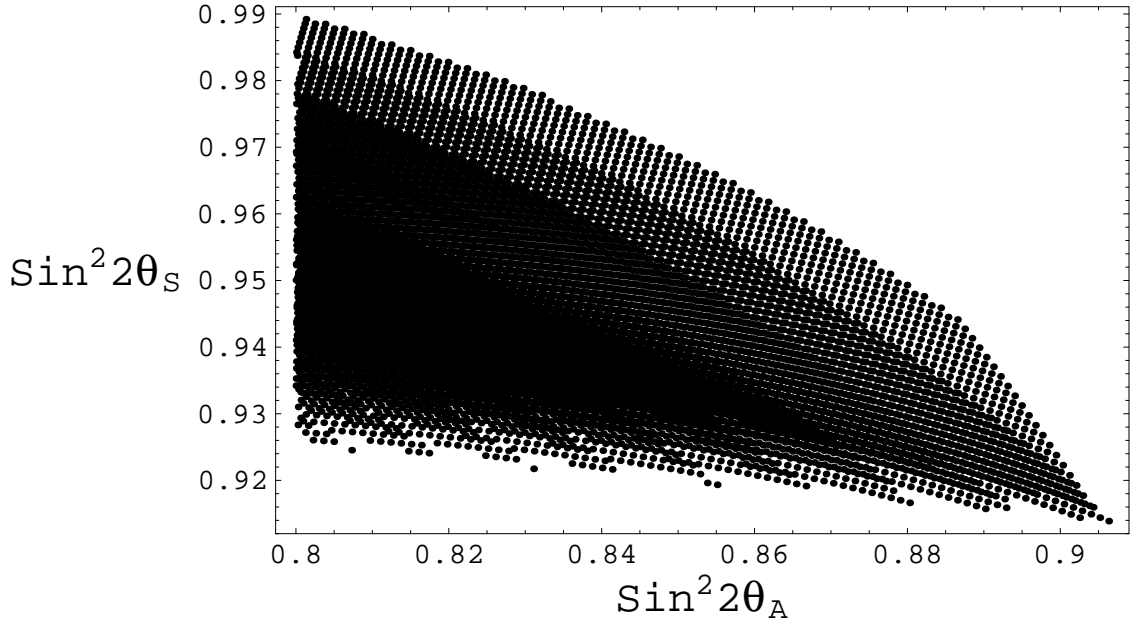}
\caption{
The figure shows the predictions for $sin^22\theta_{\odot}$ and
$sin^22\theta_A$ for the range of quark masses in table I. Note that
$sin^22\theta_{\odot}\geq 0.9$ and $sin^22\theta_A\leq 0.9$
\label{fig:cstr1}}
\end{center}
\end{figure}

\begin{figure}[h!]
\begin{center}
\epsfxsize15cm\epsffile{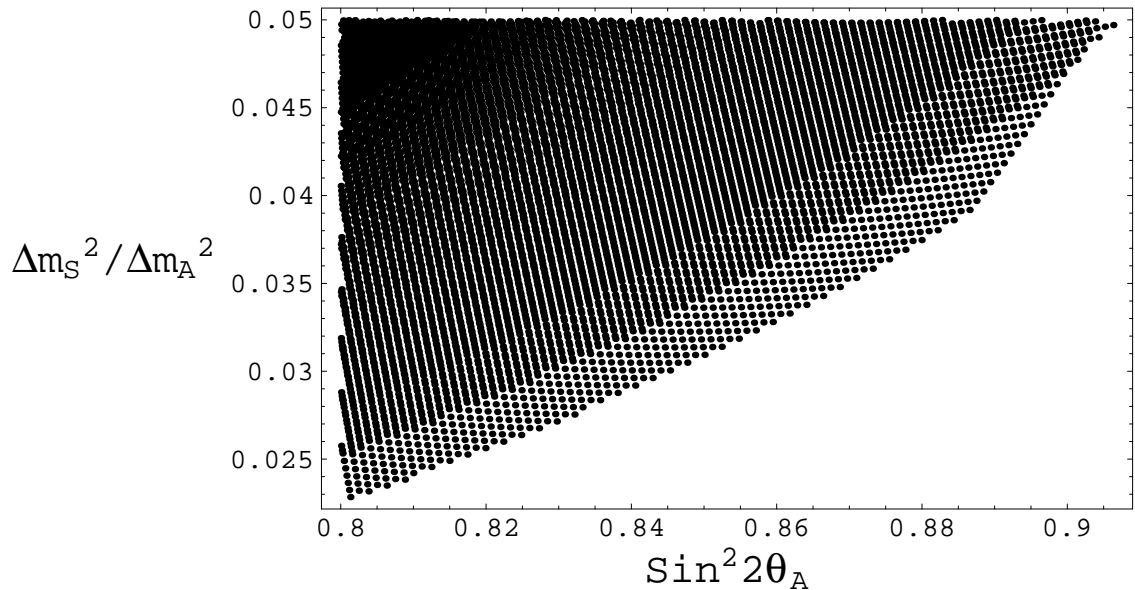}
\caption{
The figure shows the predictions for $sin^22\theta_{A}$ and
 $\Delta m^2_{\odot}/\Delta m^2_{A}$ for the range of quark masses and
mixings that fit charged lepton masses.
\label{fig:cstr2}}
\end{center}
\end{figure}

\begin{figure}[h!]
\begin{center}
\epsfxsize15cm\epsffile{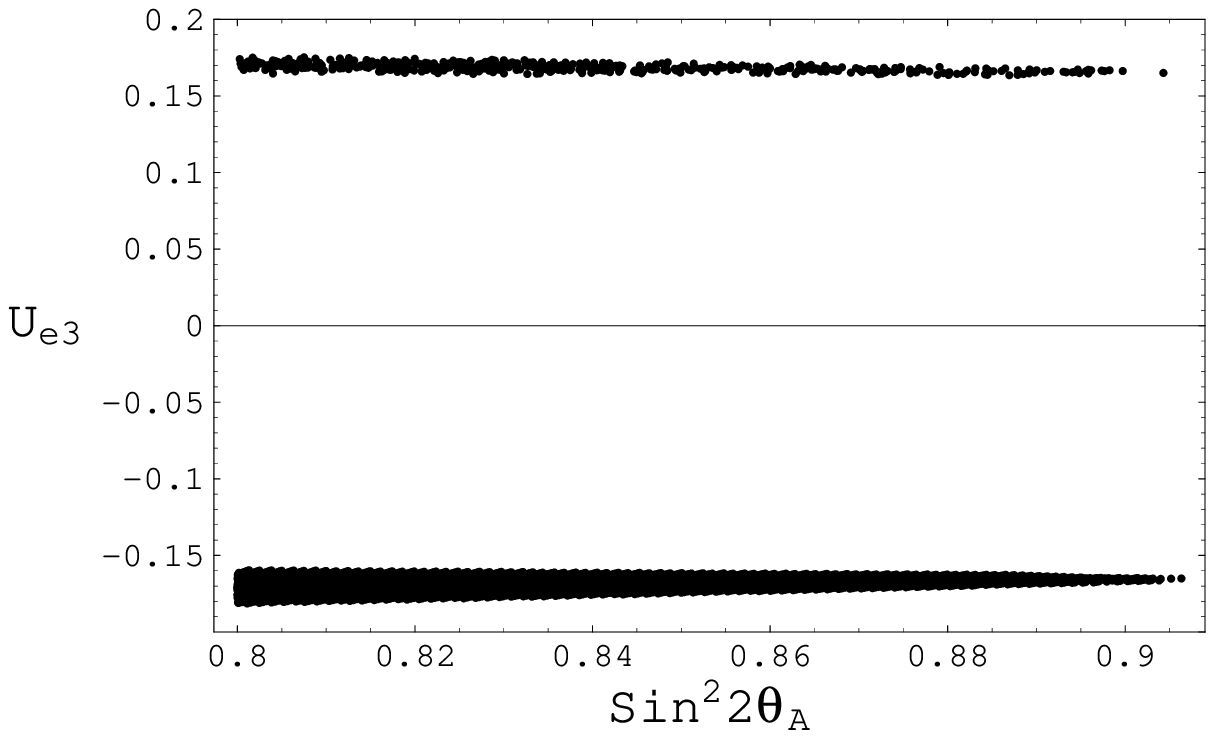}
\caption{
The figure shows the predictions of the model for $sin^22\theta_{A}$
and $U_{e3}$ for the allowed range of parameters in the model. Note that
$U_{e3}$ is very close to the upper limit allowed by the existing reactor
experiments.
\label{fig:cstr3}}
\end{center}
\end{figure}

As $tan\beta$ increases, the allowed values for the neutrino mixings and
masses fall into an even narrower range.

\section{Conclusion}
In summary, we find that a minimal SO(10) model with single {\bf 10} and
{\bf 126} Higgs coupling to matter is a completely predictive model for 
neutrino masses and can provide an excellent description of
the presently favored patterns for neutrino masses and mixings required by
data. The only assumption needed is that the $SU(2)_L$ triplet vev
dominates the
neutrino masses. No global symmetries are invoked to generate the neutrino
mass
pattern, unlike most models that employ the {\bf 16} Higgs to break the
B-L symmetry. The model predicts a hierarchical mass pattern for
neutrinos
and a value of $U_{e3}\simeq 0.16$, both of which can be tested in
upcoming long baseline neutrino experiments. The atmospheric mixing angle
is found to be between $0.8$ and $0.9$ which is also a testable prediction
of the model. In our model, the Yukawa matrices have a hierarchical
pattern, a rough understanding of which could come from introducing a
local horizontal $U(2)_H$ symmetry under which the first two families
transform as a doublet. This and other aspects of the model such as
the inclusion of a CP phase are currently under investigation. 

This work is supported by the National Science Foundation
Grant No. PHY-0099544. We thank K. S. Babu, Z. Chacko and G. Senjanovi\'c
for discussions.

\end{document}